\documentclass{article} 
\usepackage{citesort,fullpage,amsfonts,graphicx}

\newcommand{\be}{\begin{equation}}
\newcommand{\benonumber}{\begin{displaymath}}
\newcommand{\eenonumber}{\end{displaymath}}
\newcommand{\ee}{\end{equation}}
\newcommand{\bea}{\begin{eqnarray}}
\newcommand{\eea}{\end{eqnarray}}

\newcommand{\third}{\frac13}
\newcommand{\half}{\frac12}
\newcommand{\q}{\quad}
\newcommand{\eref}[1]{(\ref{#1})}
\newsymbol \gtrsim 1326
\newsymbol \lesssim 132E

\newcounter{saveeqn}%
\newcommand{\beqn}{\setcounter{saveeqn}{\value{equation}}%
	\stepcounter{saveeqn}\setcounter{equation}{0}%
	\renewcommand{\theequation}
	{\mbox{\arabic{saveeqn}\alph{equation}}}%
	\begin{eqnarray} }%

\newcommand{\eeqn}{\end{eqnarray}\setcounter{equation}{\value{saveeqn}}%
	\renewcommand{\theequation}{\arabic{equation}}}%

\newcommand{\mainlabel}[1]{\renewcommand{\theequation}%
	{\arabic{saveeqn}}\label{#1}%
	\renewcommand{\theequation}{\mbox{\arabic{saveeqn}%
	\alph{equation}}}}%

\begin{document}
\title{Interactions in Scalar Field Cosmology}
\author{\\ \\ Andrew P. Billyard\thanks{jaf@mscs.dal.ca}\\
{\small Department of Physics, Dalhousie University, Halifax, Nova Scotia, Canada, B3H  3J5}  \\ \\ \\ \\
Alan A. Coley\thanks{aac@mscs.dal.ca}\\
{\small Departments of Physics and Mathematics \& Statistics,  Dalhousie University, }\\ {\small Halifax, Nova Scotia, Canada B3H 3J5}} 
\date{}
\maketitle{

\begin{abstract}

We investigate spatially flat isotropic cosmological models which contain 
a scalar field with an exponential potential and a
perfect fluid
with a linear equation of state. We include 
an interaction term,
through which the energy of the scalar field is transferred to the
matter fields, consistent with a term that arises from
scalar--tensor theory under a conformal transformation and field
redefinition. The governing ordinary differential equations 
reduce to a dynamical system when appropriate normalized variables are
defined. We analyse the dynamical system and find that
the interaction term can significantly
affect the qualitative behaviour of the models.
The  late-time behaviour of these models may be of
cosmological interest.
In particular, for a specific range of values for the model parameters 
there are 
late-time attracting solutions, corresponding to a novel attracting equilibrium point,
which are inflationary and in which the scalar field's energy-density
remains a fixed fraction of the matter field's energy density.
These scalar field models may be of interest as late-time cosmologies, particularly
in view of the recent
observations of the current
accelerated cosmic expansion.
For appropriate values of the interaction coupling parameter,
this equilibrium point is an
attracting focus, and hence as inflating solutions approach this late-time
attractor the scalar field oscillates. 
Hence these models may also be of importance in
the study of inflation in the early universe.

\end{abstract}
}

\vspace{2in}

\noindent Pacs 98.80.Cq

\newpage 

\section{Introduction}

A variety of theories of fundamental physics predict the existence of
scalar fields \cite{Green1987a,Olive1990a,Billyard1999h}, motivating
the study of the dynamical properties of scalar fields in cosmology.
Indeed, scalar field cosmological models are of great importance in
the study of the early universe, particularly in the investigation of
inflation \cite{Guth1981a,Olive1990a}.  Recently there has also been
great interest in the late-time evolution of scalar field models.
`Quintessential' scalar field models (or slowly decaying cosmological
constant models) \cite{Caldwell1998a,Bahcall1999a} give rise to a
residual scalar field which contributes to the present energy-density
of the universe that may alleviate the dark matter problem and can
predict an effective cosmological constant which is consistent with
observations of the present accelerated cosmic expansion
\cite{Perlmutter1999a,Riess1998a}.

Models with a self-interaction potential with an exponential
dependence on the scalar field of the form
\be
V = \Lambda  e^{k \phi},
\ee
where $\Lambda$ and $k$ are positive constants, have been the subject
of much interest and arise naturally from theories of gravity
such as scalar-tensor theories or string theories
\cite{Billyard1999h}.  Recently, it has been argued that a scalar
field with an exponential potential is a strong candidate for dark
matter in spiral galaxies \cite{Guzman1998a} and is consistent with
observations of current accelerated expansion of the universe
\cite{Huterer1998a}.

A number of authors have studied scalar field cosmological models with
an exponential potential within general relativity.  Homogeneous and isotropic
Friedmann-Robertson-Walker (FRW) models were studied by Halliwell
\cite{Halliwell1987a} using phase-plane methods.
Homogeneous but anisotropic models of Bianchi types I and III (and
Kantowski-Sachs models) were studied by Burd and Barrow
\cite{Burd1988a}, Bianchi type
I models were studied by Lidsey \cite{Lidsey1992a} and Aguirregabiria {\em et
al.} \cite{Aguirregabiria1993b}, and Bianchi models of types III and VI were
studied by Feinstein and Ib\'{a}\~{n}ez \cite{Feinstein1993a}.
A qualitative analysis of Bianchi
models with $k^2<2$ (including standard matter satisfying standard
energy conditions) was completed by Kitada and Maeda
\cite{Kitada1993a}. 
The governing differential equations in spatially homogeneous Bianchi
cosmologies containing a scalar field with an exponential potential
reduce to a dynamical system when appropriate
expansion- normalized variables are defined. This dynamical system was studied in
detail in \cite{Coley1997a} (where matter terms were not considered).

One particular solution that is of great interest is the flat, isotropic
power-law inflationary solution which occurs for $k^2<2$. This
power-law inflationary solution is known to be an attractor for all initially
expanding Bianchi models (except a subclass of the Bianchi type IX
models which will recollapse) \cite{Kitada1993a,Coley1997a}. Therefore, all
of these models inflate forever; there is no exit from inflation and
no conditions for conventional reheating.

Recently cosmological models which contain both  
a scalar field with an exponential potential and a
barotropic perfect fluid
with an equation of state
\be
p = (\gamma -1)\mu,
\ee
where $\gamma$ is in the physically relevant range
$2/3<\gamma \le 2$,
have come under heavy analysis.  One class of exact solutions found for
these models has the property that the energy density due to the
scalar field is proportional to the energy density of the perfect
fluid, and hence these models have been labelled matter scaling
cosmologies \cite{Wetterich1988a,Wands1993a,Ferreira1998a}.
These matter scaling solutions 
are spatially flat isotropic models and are known to be
late-time attractors (i.e., stable) in the subclass of
flat isotropic models \cite{Wetterich1988a,Wands1993a,Ferreira1998a}
and are clearly of physical interest.
In addition to the matter scaling solutions, curvature
scaling solutions \cite{vandenHoogen1999b} 
and anisotropic scaling solutions \cite{Coley1998b}
are also possible. A comprehensive analysis of
spatially homogeneous models with a perfect fluid
and a scalar field with an exponential potential has recently been
undertaken \cite{Billyard1999f}.

Within the standard model of inflation (with, for example, a quadratic or quartic self-interaction
potential), in the early universe the microphysics of
the scalar field leads to an accelerated expansion essentially driven by the potential energy
(or vacuum energy) that arises when the scalar field is displaced from its
potential energy minimum. Provided that the potential is sufficiently flat,
the universe undergoes many e-folds of expansion and 
the matter content is driven to zero
\cite{Linde1987a,Linde1983a,Linde1983b,Amendola1996a,Berera1996a}.  
As the scalar field nears its minimum,
the vacuum energy is converted to coherent oscillations of the scalar field 
(which corresponds to non-relativistic scalar particles). Eventually,
since the scalar field is coupled
to other (fermionic and bosonic) matter fields, these
particles decay into lighter particles and their thermalization results in the
`reheating' of the universe (thereby accounting for the current entropy
in the universe today).

Although the exponential models are interesting models for a variety
of reasons, they have some shortcomings as inflationary models.
While Bianchi models generically asymptote towards
the power-law inflationary model in which the matter terms are driven to
zero for $k^2<2$,
there is no graceful exit from this inflationary phase.
Furthermore, the scalar field cannot oscillate 
and so reheating cannot occur by the conventional scenario.
Clearly, these models need to be augmented in an attempt to alleviate
these problems.  For example, exponential potentials are only believed to
be an approximation, and so the theory could include more complicated
potentials (although it is not clear what these other potentials should be).

The goal of this paper is to examine how interaction terms,
through which the energy of the scalar field is transferred to the
matter fields, can
affect the qualitative behaviour of models containing matter and a
scalar field with an exponential potential.  Such an interaction term,
denoted by $\delta$, arises in the conservation
equations, via
\beqn
\mainlabel{conserve_interact}
\label{phi_con}
&& \dot\phi\left(\ddot\phi +3H\dot\phi +kV\right) = -\delta \\
\label{mu_con}
&& \dot\mu + 3\gamma H\mu = \delta.
\eeqn
The form of such a term has been discussed in the literature within
the context of inflation and reheating. 
An alternative to the conventional reheating model,
in which the
scalar field's energy is transferred to the matter
due to scalar field oscillations,
is the warm
inflationary model \cite{Berera1995a,Bellini1999a,Ferreira1998a}
in which an interaction term is significant throughout the inflationary
regime (not just after slow-roll) and so the energy of the scalar
field is continually transferred to the matter content throughout inflation
and the matter content is {\em not} driven to zero.

Several examples of interaction terms appear in the literature for
models with a variety of self-interaction
potentials. In particular,  potentials which have a global
minimum have attracted much attention.  For example, Albrecht {\em et al.} \cite{Albrecht1982b}
considered $\delta=a\dot\phi^d\phi^{5-2d}$ (where $a$ is a constant),
derived from dimensional arguments, in the reheating context after
inflation (with potentials derived from Georgi--Glashow SU(5) models).
In a similar context, Berera
\cite{Berera1995b} considered interaction terms of the form $\delta =
a\dot\phi^2$.  Quadratic potentials and interaction terms of the form
$\delta= a\dot\phi^2$ and $\delta= a\phi^2\dot\phi^2$ were considered
by de Oliveira and Ramos \cite{Oliveira1998a} and a
graceful exit from inflation was demonstrated numerically. Similarly, Yokoyama {\em et al.}
\cite{Yokoyama1987a}
showed that an interaction term, which is 
negligible during the slow-roll inflationary phase, dominates at
the end of inflation when the scalar field is oscillating about its
minimum; during this reheating phase it is assumed that the energy
transferred from the scalar field is solely converted into particles.

Within the context of exponential potentials, Yokoyama and Maeda
\cite{Yokoyama1988a} and Wands {\em et al.}
\cite{Wands1993a} considered interactions of the form
$\delta=a\sqrt{V}$.  The main goal in both these papers was to show
that power-law inflation can occur for $k^2>2$, thereby showing that
inflation can exist for steeper potentials.  The main motivation for this 
work is the fact that exponential
potentials which arise naturally from other theories, such as
supergravity or superstring models, typically have $k^2>2$.
Wetterich
\cite{Wetterich1995a} considered interaction terms containing a matter
dependence, namely $\delta = a \dot\phi\mu$, in which perturbation analysis
showed that the matter scaling solutions were stable solutions when
such interaction terms are included.  In \cite{Wetterich1995a}, it was
shown that the age of the universe is older when $\delta$ is included
and that the scalar field can still significantly contribute to the
energy density of models at late times.  In \cite{Billyard1999h}, certain string
theories in which the energy sources are separately conserved in the
Jordan frame naturally lead to interaction terms in the Einstein
frame, although this is {\em not} specific to string cosmologies; any
scalar-tensor theory with matter terms and a power-law potential will
yield the same results \cite{Wetterich1995a}.

The scalar field in scalar--tensor theory can be related to the scalar
field in the general relativistic scalar field theory via a conformal
transformation.  If the matter terms are separately conserved in the
scalar--tensor theory (Jordan frame) then they lead to the following
conservation equation in the Einstein Frame
\begin{equation}
\label{couplingfromST}
\dot\phi\left(\ddot\phi +3H\dot\phi +kV\right) = 
\frac{\left(4-3\gamma\right)}{2\sqrt{\omega+\frac{3}{2}}} \dot\phi \mu
\end{equation}
(and a similar equation for $\dot\mu$), where $\omega$ is the coupling
parameter in the scalar--tensor theory.  This therefore leads to an
interaction term of the form $\delta=-a\dot\phi\mu$, where
\begin{equation}
a = \frac{\left(4-3\gamma\right)}{2\sqrt{\omega+\frac{3}{2}}}.
\end{equation}
Similarly, in string theory ($\omega=-1$), a similar interaction term
arises in the Einstein frame which depends on the energy density of
the matter (axion) field and $a$ is of order unity.  Finally, a term
of the form $\delta=a\mu H$ might be motivated by analogy with
dissipation.  For example, a fluid with bulk viscosity may give rise
to a term of this form in the conservation equation \cite{Eckart1940a}.

If $\delta$ does not depend on the matter energy-density, then
unphysical situations may arise.  For conventional interaction terms
without $\mu$, we have found numerically that $\mu$ is driven to zero
in a finite time and subsequently the matter energy-density becomes
negative (see \cite{Billyard1999h} for details).  Thus, we shall
include a factor $\mu$ in the interaction term in order to ensure
that $\mu\geq 0$.  In
\cite{Wetterich1995a} it was shown that if $\delta<0$ at the equilibrium
points then static solutions (i.e. $\dot\phi=0$) are {\em not} possible.
Furthermore, we require that the sign of $\delta$ be positive at
equilibrium points representing inflationary phases, otherwise the
matter fields will be ``feeding'' the scalar field and will
redshift to zero even faster than in the absence of the interaction
terms.

This paper examines interaction terms of the general form
$\delta=\bar\delta\mu H$ (where
$\bar\delta=\bar\delta(\dot\phi,V,H)$) in the context
of flat FRW models with emphasis in determining the asymptotic properties of
these models.  In particular, it will be determined whether these
models can asymptote towards inflationary models in which the matter
terms are not driven to zero.  This would partially alleviate the need
for reheating; since the matter content tracks that of scalar field
(in this context) it is never driven to zero (unless both are driven
to zero in which case the solution is not inflationary).  However, a
more comprehensive reheating model would still be necessary.  The
structure of the paper is as follows.  In section
\ref{i_govern}, the governing equations are defined and the case $\delta=0$  
studied in \cite{Copeland1997a} is reviewed.  In section \ref{i_I}, the case
$\delta=a\dot\phi\mu$ is studied, motivated by the conformal relationships
between the Jordan and Einstein frame in string theory, and extends
the work of \cite{Wetterich1995a}.  In section \ref{ii_II} we shall discuss
the results of the analysis and in 
section \ref{i_II} we shall examine an interaction term
of the form
$\delta=a\mu H$.
The paper ends with conclusions in section \ref{i_discuss}.

\section{Governing Equations\label{i_govern}}

The governing field equations are given by the equations (\ref{conserve_interact}) and
\be 
\label{i_Hdot}
\dot{H} = -\frac{1}{2} (\gamma \mu + \dot{\phi}^2), 
\ee
subject to the Friedmann constraint
\be
\label{i_Fried_1}
H^2 = \frac{1}{3} (\mu + \frac{1}{2} \dot{\phi}^2 + V), 
\ee
(an overdot denotes ordinary
differentiation with respect to time $t$).  Note that the
total energy density of the scalar field is given by
$\mu_\phi = \frac{1}{2}\dot{\phi}^2 +V$. 
The deceleration parameter for this system is given by
\be
q = \third H^{-2} \left[\dot\phi^2-V+\half\left(3\gamma-2\right)\mu \right],
\ee
and is {\em independent} of the interaction term.

Defining
\be
\label{i_def_eqns}
x \equiv \frac{\dot{\phi}}{\sqrt{6}H} \q , \q y \equiv 
	\frac{\sqrt{V}}{\sqrt{3}H},
\ee
and the new logarithmic time variable $\tau$ by
\be
\label{i_new_time}
\frac{d \tau}{dt} \equiv H,
\ee
the governing differential equations can be written as the
plane-autonomous system:
\beqn
\mainlabel{i_xy_1}
\label{i_x_1}
x' & =& -3x - \sqrt{\frac{3}{2}} k y^2 + \frac{3}{2} x [2x^2 + \gamma (1-x^2 -y^2)] - \frac{\bar\delta}{6x}(1-x^2 -y^2), \\
\label{i_y_1}
y' &=& \frac{3}{2}y \left[\sqrt{\frac{2}{3}} k x + 2x^2 + \gamma (1-x^2 -y^2)\right], 
\eeqn
where a prime denotes differentiation with respect to $\tau$ \cite{Billyard1999h}.
Note that $y=0$ is an invariant set, corresponding to $V=0$.  The
equations are invariant under $y\rightarrow -y$ and $t\rightarrow -t$
and so the region $y<0$ is a time-reversed mirror of the region $y>0$;
therefore, only $y>0$ will be considered.  Similarly, only $k>0$ will be
considered since the equations for the interaction terms to be considered are invariant under $k\rightarrow -k$ and
$x\rightarrow -x$.

Equation (\ref{i_Fried_1}) can be written as 
\benonumber \Omega + \Omega_\phi =1, \eenonumber
where
\be
\label{i_O_1}
\Omega \equiv \frac{\mu_\gamma}{3 H^2}, \q \Omega_\phi \equiv 
	\frac{\mu_\phi}{3H^2} = x^2 + y^2, 
\ee
which implies that $0 \leq x^2 +y^2 \leq 1$ for $\Omega \geq 0$ so that the
phase-space is bounded.  The deceleration parameter is now written
\be
q = -1 +3 x^2+\frac{3}{2}\gamma\left(1-x^2-y^2\right).
\ee

\subsection{Comments on arbitrary $\bar\delta$}

The fact that $q$ is independent of the interaction term implies that
the region of phase space which represents inflationary models is the
same for all of the models considered.  Namely, $q=0$ occurs along the
ellipse $\gamma y^2 = (2-\gamma)x^2+\third(3\gamma-2)$.  For any value
of $\gamma$, the lines intersect the boundary of the phase space at
$x^2=\third$.

It is possible to make some qualitative comments about the system
(\ref{i_xy_1}) for arbitrary $\bar\delta$.  First, {\em the location of equilibrium points} on the
boundary $x^2+y^2=1$ ($\Omega=0$) are independent of the choice of
such interaction terms;  the three equilibrium points (and their
associated eigenvalues) which exist on the boundary for any $\bar\delta$ are:
\beqn
\mainlabel{boundary_points}
{\cal K}^+:&& (x,y)=(+1,0) \\ \nonumber
&&	\left(\lambda_1,\lambda_2\right) = \left(3(2-\gamma)+\third 
	\left.\bar\delta\right|_{{\cal K}^+},\sqrt{\frac{3}{2}}
	\left[\sqrt6 + k\right]\right), \\
{\cal K}^-:&& (x,y)=(-1,0) \\ \nonumber
&&	\left(\lambda_1,\lambda_2\right) = \left(3(2-\gamma)+\third 
	\left.\bar\delta\right|_{{\cal K}^-}, \sqrt{\frac{3}{2}}
	\left[\sqrt6 - k\right]\right), \\
P_{\cal S}:&& (x,y)=\left(-\frac{k}{\sqrt6},\sqrt{1-k^2/6} \right) \\ \nonumber
&&	\left(\lambda_1,\lambda_2\right) = \left(-\half\left[6-k^2\right],
	-\left[3\gamma-k^2-\third\left.\bar\delta\right|_{P_{\cal S}}\right]
	\right).
\eeqn
The points ${\cal K}^\pm$ represent the isotropic subcases of Jacobs'
Bianchi type I solutions \cite{Collins1971a} (subcases of the Kasner
models), generalized to include a massless scalar field.  These
solutions are non-inflationary ($q=2$).  The point $P_{\cal S}$, which
exists only for $k^2<6$, represents the FRW power-law model
\cite{Halliwell1987a,Coley1997a} and is inflationary for $k^2<2$
($q=\half\left[k^2-2\right]$).  Although these three points exists for
any $\bar\delta$, the interaction term {\em does} affect the stability
of these solutions, as is evident from the eigenvalues in \eref{boundary_points}.  In
particular, for $k^2<3\gamma$ the point $P_{\cal S}$ can become a
saddle point if
\benonumber
\left.\bar\delta\right|_{P_{\cal S}} > 3\left(3\gamma-k^2 \right).
\eenonumber
Hence, if the interaction term is significant, solutions will spend an
indefinite period of time near this power-law inflationary model, but
will then evolve away and typically be attracted to another
equilibrium point in some other region of the phase space.

Matter scaling solutions (i.e. those solutions in which
$\gamma_\phi=\gamma$), denoted by ${\cal F}_{\cal S}$ in 
\cite{Billyard1999f}, exist only in special circumstances when such
interaction terms are present, and occur at the point
\be
x_{{\cal F}_{\cal S}} = -\sqrt\frac{3}{2}\frac{\gamma}{k}, \qquad 
	y_{{\cal F}_{\cal S}} = \sqrt{\frac{3\gamma(2-\gamma)}{2k^2}}.
\ee
Substituting these solutions into equations (\ref{i_xy_1}) yields
\beqn
x' &=& \frac{(3\gamma-k^2)}{3\sqrt6k\gamma}
	\left.\bar\delta\right|_{{\cal F}_{\cal S}}, \\
y' &=& 0.
\eeqn
Hence, the matter scaling solutions will be represented by an
equilibrium point only if $\left.\bar\delta\right|_{{\cal F}_{\cal
S}}=0$ (or in the special case $3\gamma=k^2$, which is typically a
bifurcation value).  For the simple forms for $\delta$ given in the literature
and those used in this paper, this condition will not be satisfied
and so the matter scaling solutions cannot be asymptotic attracting
solutions.  

However, an analogous situation does arise.  In
particular, any equilibrium point within the boundary of the phase
space will satisfy
\be
y_0^2 = \frac{(2-\gamma_\phi)}{\gamma_\phi} x_0^2,
\ee
in which the scalar field is equivalent to a perfect fluid of the form
$p_\phi=(\gamma_\phi -1)\mu_\phi$, but where $\gamma_\phi\neq\gamma$.
Consequently, any attracting equilibrium point within the phase space
will represent models in which neither the matter field nor the scalar
field is negligible and the scalar field mimics a barotropic fluid
different from the matter field and therefore could still constitute a
possible dark matter candidate.

Finally, it can be shown that any equilibrium point within (but not
on) the boundary will occur for $x_0<0$.  For $y\neq0$ and $\Omega\neq 0$ equation (\ref{i_y_1}), which does not depend on $\delta$, yields 
\be
\label{yp_zero}
\gamma y^2 = \gamma +\sqrt\frac{2}{3}kx +(2-\gamma)x^2,
\ee
a relationship any such equilibrium point must satisfy. Now, $\gamma(y^2+x^2)<\gamma$ since $y^2+x^2<1$ inside the boundary, and hence  (\ref{yp_zero}) yields
\be
x\left(x+\frac{k}{\sqrt 6}\right) <0,
\ee
which cannot be satisfied for $x>0$ (since $k>0$).

\subsection{Review of the case $\delta=0$\label{i_d0}}

Copeland {\em et al.} \cite{Copeland1997a} performed a phase-plane
analysis of the system (\ref{i_xy_1}) for $\delta=0$, and found five
equilibrium points.  One of the equilibrium points (denoted here by
$P$) represents a flat, non-inflating FRW model
\cite{Copeland1997a}, for which $\Omega=1$.   For $2/3<\gamma<2$ this point is a
saddle in the phase space.  The flat FRW matter scaling solution
 ($F_{\cal S}$) was found to exist for $k^2>3\gamma$ and was shown to be a
 sink.  The equilibrium point ${\cal K}^+$ was shown to be a
 source for all $k$ and ${\cal K}^-$ a source for $k^2<6$.  The FRW
 power-law model ($P_{\cal S}$) was shown to be a sink
 for $k^2<3\gamma$, and was shown to represent an inflationary model for $k^2<2$.  The results found in \cite{Copeland1997a} are
 summarized in table \ref{i_table1}.
\begin{table}[ht]
\begin{center}
\begin{tabular}{|l||c|c|c|c|}
\hline
	& $0\leq k^2 \leq 2$ 	& \multicolumn{2}{|c|}{$2<k^2<6$} 		& $k^2>6$ \\ \hline 
	&  	& $k^2<3\gamma$ & $k^2>3\gamma$ &  \\ \hline 
\hline
$P$ 	& {\scriptsize saddle}	& \multicolumn{2}{|c|}{\scriptsize saddle}	& {\scriptsize saddle} \\
	& {\tiny (NI)}		& \multicolumn{2}{|c|}{\tiny (NI)}		& {\tiny (NI)} \\
\hline
${\cal K}^+$ & source 		& \multicolumn{2}{|c|}{source} 	& source \\
	& {\tiny (NI)}		& \multicolumn{2}{|c|}{\tiny (NI)}		& {\tiny (NI)} \\
\hline
${\cal K}^-$ & source 		& \multicolumn{2}{|c|}{source} 	& {\scriptsize saddle} \\
	& {\tiny (NI)}		& \multicolumn{2}{|c|}{\tiny (NI)}		& {\tiny (NI)} \\
\hline
$P_{\cal S}$ & sink 	&  sink &  {\scriptsize saddle} & \tiny{DNE} \\
	& {\tiny (I) (for $k^2<2$)} & {\tiny (NI)} & {\tiny (NI)}&  \\
\hline
${\cal F}_{\cal S}$ & {\tiny DNE} &{\tiny DNE}	&  sink & sink \\
	& && {\tiny (NI)}	&  {\tiny (NI)} \\
\hline
\end{tabular}
\end{center}
\caption[Equilibrium points for $\delta=0$]{{\em The equilibrium
points for $\delta=0$ and their stability for various values of $k$.
The label ``(NI)'' denotes non-inflationary models whereas ``(I)''
represents inflationary models, and ``DNE'' is used when an
equilibrium point does not exist.}\label{i_table1}}
\end{table}

\section{Interaction term of the form $\delta=-a\dot\phi\mu$\label{i_I}}

In this section, an interaction term of the form
$\delta=-a\dot\phi\mu$ (and hence $\bar\delta=-a\dot\phi/H$) shall be considered.  Again, it will be assumed that $k>0$.  The explicit sign choice for $\delta$, with the
assumption that $a>0$, is to guarantee that all equilibrium points
within the phase space will represent models in which energy is being
transferred from the scalar field to the perfect fluid, since it was
shown that all equilibrium points within the phase space occur for
$x<0$ ($\dot\phi<0$).  Indeed, this is even true for the equilibrium
point $P_{\cal S}$ on the boundary $\Omega=0$, since it is located at
$x<0$.  With this particular choice for $\delta$, equations
(\ref{i_xy_1}) become
\beqn
\mainlabel{i_xy_2}
\label{i_x_2}
x' & =& -3x(1-x^2) - \sqrt{\frac{3}{2}} k y^2 + \left(\frac{3}{2}\gamma x 
	+ \sqrt{\frac{3}{2}}a\right)\left(1-x^2-y^2\right), \\
\label{i_y_2}
y' &=& \frac{3}{2}y \left[\sqrt{\frac{2}{3}} k x + 2x^2 + 
	\gamma (1-x^2 -y^2)\right].
\eeqn

\noindent There are five equilibrium points for this system:
\begin{enumerate}
\item ${\cal K}^+:\qquad (x,y)=(+1,0), \qquad \Omega=0, \qquad q=2.$ \\
The eigenvalues for this equilibrium point are 
\be
\left(\lambda_1,\lambda_2\right) = \left(3(2-\gamma)-\sqrt6 a,
		\sqrt{\frac{3}{2}}\left[\sqrt6 + k\right]\right)
\ee
This equilibrium point is a source for $a<\sqrt\frac{3}{2}(2-\gamma)$
and a saddle otherwise.

\item ${\cal K}^-:\qquad (x,y)=(-1,0), \qquad \Omega=0, \qquad q=2.$ \\
The eigenvalues for this equilibrium point are 
\be
\left(\lambda_1,\lambda_2\right) = \left(3(2-\gamma)+\sqrt6 a, 
	\sqrt{\frac{3}{2}} \left[\sqrt6 - k\right]\right),
\ee
and so ${\cal K}^-$ is a source for $k^2<6$ and a saddle otherwise.

\item $\displaystyle P_{\cal S}:\qquad
(x,y)=\left(-\frac{k}{\sqrt6},\sqrt{1-\frac{k^2}{6}} \right), \qquad \Omega=0,
\qquad q=\half(k^2-2).$  \\ 
The eigenvalues for this equilibrium point are 
\be
\left(\lambda_1,\lambda_2\right) = \left(-\half\left[6-k^2\right],
	-\left[3\gamma-k^2-ka\right]\right).
\ee
This point exists only for $k^2<6$ (when $k^2=6$, $P_{\cal S}$
merges with the equilibrium point ${\cal K}^-$).  Here, 
$P_{\cal S}$ is a sink for $a<(3\gamma-k^2)/k$ and a saddle
otherwise.

\item $\displaystyle N:\qquad (x,y)=\left(-\sqrt\frac{3}{2}
	\frac{\gamma}{\Delta},\sqrt{\frac{3\gamma(2-\gamma)+2a\Delta}
	{2\Delta^2}} \right), \qquad
	\Omega=\frac{k\Delta-3\gamma}{\Delta^2},$ 
\qquad
$\displaystyle q=\frac{3\gamma k-2\Delta}{\Delta},$ 
	\hspace*{4em}\\
where $\Delta\equiv k+a>0$.  Note that this solution is physical
(i.e., $\Omega\geq 0)$ either for $k^2>3\gamma$ or for $k^2<3\gamma$  and
\be
a\geq (3\gamma-k^2)/k\label{exist_1}.
\ee  
These solutions were discussed in \cite{Wetterich1995a} for $a<k$ and
are related to similar power-law solutions discussed in
\cite{Wands1993a}.  This model inflates if
\be 
a\geq(\frac{3}{2}\gamma-1)k\label{inflate_1}.
\ee
(since only $a<k$ was considered in \cite{Wetterich1995a}
the solutions therein were {\em not} inflationary).  For
$k^2<2$, if condition (\ref{exist_1}) is satisfied then
(\ref{inflate_1}) is automatically satisfied and so these models
inflate for $k^2<2$.  For $2<k^2<3\gamma$, if condition
(\ref{inflate_1}) is satisfied then (\ref{exist_1}) is automatically
satisfied and therefore models can inflate for $k^2>2$ if
$a\geq(\frac{3}{2}\gamma-1)k$.  For $k^2>3\gamma$ there is no constraint on
$a$ for the point to exist and therefore whether this models inflates
is solely determined by (\ref{inflate_1}).  The eigenvalues for this
equilibrium point are \index{exact solutions!scaling-like solutions ($N$)}
\begin{eqnarray}
\label{newsink_1}
\lambda_\pm &=& \frac{-3\left[(2-\gamma)k+2a\right]}{4\Delta}
\pm \frac{\sqrt{9\left[(2-\gamma)k+2a\right]^2 
	-24\left[3\gamma(2-\gamma)+2a\Delta\right] \left[k\Delta-3\gamma\right]
		}}{4\Delta},
\end{eqnarray}
and so $N$ is always a sink when it exists.  Note that the scalar
field acts as a perfect fluid with an equation of state parameter given by
\be
\gamma_\phi = \frac{\gamma}{1+\frac{a\Delta}{3\gamma}} < \gamma.
\ee

\item $\displaystyle N_2:\qquad (x,y)=\left(\sqrt\frac{2}{3}
	\frac{a}{(2-\gamma)},0\right), \qquad
	\Omega=1-\frac{2a^2}{3(2-\gamma^2)},$ 
\quad
$\displaystyle q=\half(3\gamma-2)
	+\frac{a^2}{(2-\gamma)}>0.$\hspace*{5em}\\
This equilibrium exists for $a < \sqrt\frac{3}{2}(2-\gamma)$ and is a saddle, as determined from its eigenvalues:
\be
\left(\lambda_1,\lambda_2\right) = \left(-\frac{3}{2}[2-\gamma]
	\left[1-\frac{2a^2}{3(2-\gamma)^2}\right], \frac{3}{2}\gamma+
	\frac{a(k+a)}{(2-\gamma)}\right).
\ee
\end{enumerate}

Table \ref{i_table2} lists the equilibrium points and their stability
for the ranges of $k$ and $a$.
As is evident, the presence of the interaction term can substantially
change the dynamics of these models.  We note that all equilibrium points
correspond to self--similar models \cite{Carr1999a}.

\begin{table}[ht]
\begin{center}
\begin{scriptsize}
\begin{tabular}{|l||c|c||c|c|c||c|c||c|c|}
\hline
& \multicolumn{2}{|c||}{$0< k^2 < 2$} & \multicolumn{3}{|c||}{$2<k^2<3\gamma$}
& \multicolumn{2}{|c||}{$3\gamma<k^2<6$} &\multicolumn{2}{|c|}{$k^2>6$}\\ 
\hline 
& $a<C_1$ & $a>C_1$ & $a<C_1$ &$C_1<a<C_2$ & $a>C_2$&$a<C_2$&$a>C_2$&$a<C_2$ 
& $a>C_2$
\\ \hline 
\hline
${\cal K}^+$ 	& \multicolumn{9}{|c|}{\rule[0em]{0em}{1.2em}{\bf R} 
			for $2a^2<3(2-\gamma)^2$} \\ 
   {\tiny(NI)}	& \multicolumn{9}{|c|}{s for $2a^2>3(2-\gamma)^2$ } \\ \hline 
${\cal K}^-$ 	& \multicolumn{7}{|c||}{\rule[0em]{0em}{1.2em}
			\raisebox{-.7em}[0pt]{\bf R}}&\multicolumn{2}{|c|}
			{\raisebox{-.7em}[0pt]{s}} \\ 
   {\tiny(NI)}	&  \multicolumn{7}{|c||}{ } &\multicolumn{2}{|c|}{ }\\ \hline 
$P_{\cal S}$ 	&  \rule[0em]{0em}{1.2em} {\bf A} & s & {\bf A} 
			& \multicolumn{4}{|c||}{s} &\multicolumn{2}{|c|}
			{\raisebox{-.7em}[0pt]{\tiny DNE}} \\
	& {\tiny (I)} & {\tiny (I)}& {\tiny (NI)} 
		 &\multicolumn{4}{|c||}{{\tiny(NI)}}&\multicolumn{2}{|c|}{ } \\
		\hline
$N$ 		 & \rule[0em]{0em}{1.2em}\raisebox{-.7em}[0pt]{\tiny DNE} & {\bf A} 
			& \raisebox{-.7em}[0pt]{\tiny DNE} & {\bf A}& {\bf A}
			&{\bf A}&{\bf A}&{\bf A}&{\bf A}\\
   	&     	 & {\tiny (I)} &   & {{\tiny (NI)}} & {{\tiny (I)}} 
			&{{\tiny (NI)}} &{{\tiny (I)}} & {{\tiny (NI)}}
			&{{\tiny (I)}}\\\hline
$N_2$ 		 & \multicolumn{9}{|c|}{\rule[0em]{0em}{1.2em} s for 
			$2a^2<3(2-\gamma)^2$} \\ 
     {\tiny(NI)} & \multicolumn{9}{|c|}{DNE for $2a^2>3(2-\gamma)^2$}\\ \hline 
\end{tabular}
\end{scriptsize}
\end{center}
\caption[Equilibrium points for $\delta=-a\dot\phi\mu$]{{\em The equilibrium
points for the model with $\delta=-a\dot\phi\mu$ and their stability
for various values of $k$ and $a$.  Note that
$C_1\equiv(3\gamma-k^2)/k$ and $C_2\equiv(3\gamma-2)k/2$.  The symbol
``{\bf R}'' denotes when the equilibrium point is a source (repellor),
``s'' for when it is a saddle, ``{\bf A}'' for when it is a sink
(attractor), and ``DNE'' when it does not exist within the particular
parameter space.  The label ``(NI)'' denotes non-inflationary models
whereas ``(I)'' represents inflationary models.}\label{i_table2}}
\end{table}

\section{Discussion\label{ii_II}}

\subsection{Inflation}

We first note that we can obtain inflationary solutions when $k^2>2$, unlike
the case in which there is no interaction term. Moreover, we see from Table 2
that these inflationary solutions, corresponding to the equilibrium point $N$
and which occur for $a> \half(3\gamma-2)k$,
are sinks (attractors). This result complements the results of
\cite{Wands1993a,Yokoyama1988a} who looked for inflationary solutions
for steeper potentials. 
When $k^2<2$ and $0<a<(3\gamma-k^2)/2$ the power-law inflationary solution
corresponding to the equilibrium point $P_{\cal S}$
is again a global attracting solution.
 
Of particular interest is the case when $k^2<2$ and
$a>(3\gamma-k^2)/k>0$, when $P_{\cal S}$ is no longer a sink.
Therefore, trajectories approach this equilibrium point (and the
models inflate for a definite but arbitrarily large period of time)
and eventually asymptote towards the {\em new} inflating model
corresponding to $N$.  For $a\approx (3\gamma-k^2)/k$, the eigenvalues
for $N$ (\ref{newsink_1}) are real and negative and so the attracting
solution is represented by an attracting node.  However, for
$a\gtrsim(3\gamma-k^2)/2$ this equilibrium point is a {\em spiral
node}; i.e., trajectories exhibit a {\em decaying oscillatory
behaviour} as they asymptote towards $N$. (For example, for large $a$
equation (\ref{newsink_1}) becomes $\lambda_\pm \approx -\frac{3}{2}
\pm \sqrt{-3ak}$, leading to complex eigenvalues.)

\

\

\noindent
{\em An Example}

To illustrate this oscillatory nature, an explicit example is chosen
with $k=1$ and $\gamma=4/3$ (radiation).  The equilibrium points
and their respective eigenvalues are:
\beqn
{\cal K}^\pm: && \left(\lambda_1,\lambda_2\right) = \left(\pm\sqrt6 a+2,
	\sqrt\frac{3}{2}\right),\\\nonumber\\
P_{\cal S}: &&\left(\lambda_1,\lambda_2\right) = \left(a-3,-\frac{5}{2}\right),\\
\nonumber \\
N: && (x,y)=\lambda_\pm =\frac{-(3a+1)}{2(a+1)}\pm \frac{\sqrt{49+26a
	+33a^2-12a^3}}{2(a+1)},
\eeqn
where $a>3$ in order for $N$ to exist and be a sink and for $P_{\cal
S}$ to be a saddle.  Note that for $a>3$, $N_2$ does not exist, ${\cal
K}^+$ is a saddle and ${\cal K}^-$ is a source.  Numerical analysis
shows that $N$ is a spiral source for $a\gtrsim 3.65$.  Figure 1
depicts this phase space for a typical value of $a$ in this range (for
illustrative purposes the value $a=8$ is taken), and the attracting
region therein is magnified in figure 2. These figures are typical for
other values of $\gamma$ (this comment is important since we note
that in the context of conformally transformed scalar-tensor theories,
strictly speaking $\delta=0$ for $\gamma=4/3$).

\begin{figure}[htp]
  \centering
   \includegraphics*[width=5in]{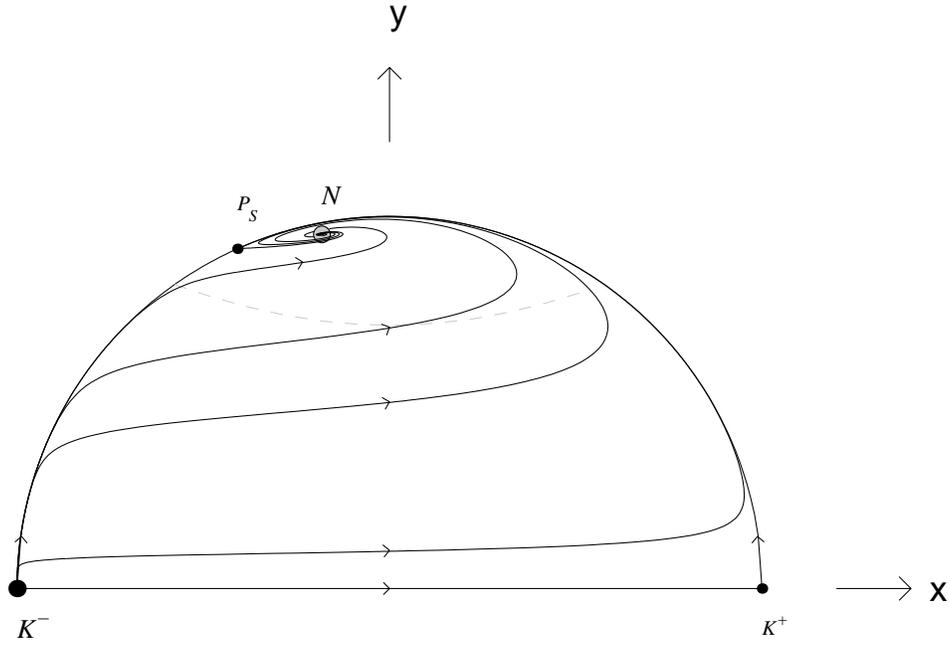}
  \caption{{\em Phase diagram of the system
(\ref{i_xy_1}) when $\delta=-a\dot\phi\mu$ for the choice of parameters
$k=1$, $\gamma=4/3$ and $a=8$.  In this figure, the black dot
represents the source (i.e., the point ${\cal K}^-$), the large grey dot
represents the sink (i.e., the point $N$) and small black dots
represent saddle points.  The region above the grey dashed line
represents the inflationary portion of the phase space. Arrows on the
trajectory indicate the direction of time. }}
\end{figure}

\begin{figure}[htp]
  \centering
   \framebox{\includegraphics*[width=5in]{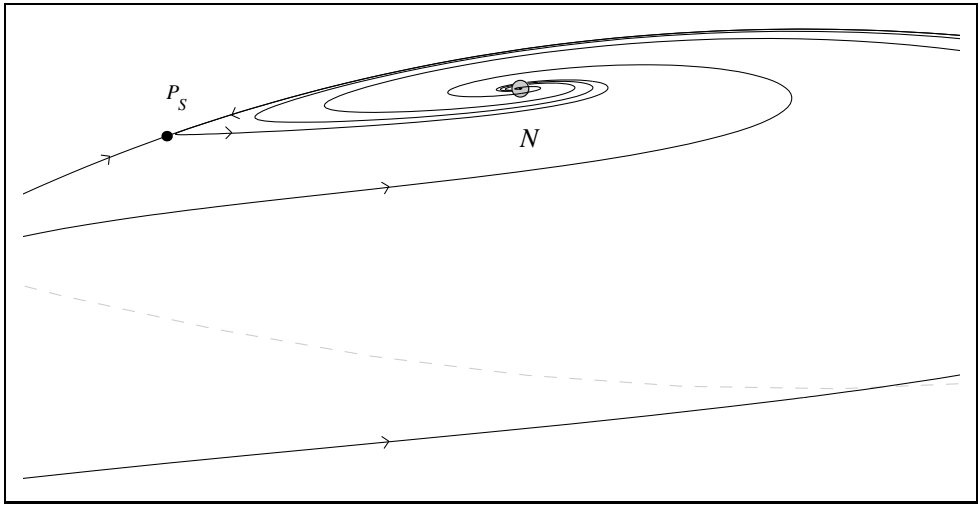}}
  \caption{{\em A magnification of the attracting region of
the phase space depicted in figure 1.  See also caption
to figure 1.}}
\end{figure}

\subsection{Late-time behaviour}

The  late-time behaviour of these models, both inflationary and non-inflationary, may also be of
cosmological interest.
For $2<k^2<3\gamma$ \& $a<(3\gamma-k^2)/k$ the late-time attracting
equilibrium point is  $P_{\cal S}$ which represents a
power-law
non-inflationary model.
Due to recent observations of 
accelerated expansion  \cite{Perlmutter1999a,Riess1998a}, models that are presently
inflating are also of interest.
For $k^2<2$ \& $a<(3\gamma-k^2)/k$ the late-time attracting
equilibrium point is $P_{\cal S}$ which represents a power-law
inflationary model.
However, in both of these cases the matter contribution is
negligible and so these models are not of physical interest.
For $k^2<2$ \& $a>(3\gamma-k^2)/k$ the late-time attracting
equilibrium point is $N$, which represents a power-law
inflationary model in which both the matter and the scalar field are
non-negligible and their energy densities are proportional to one another.
These models are potentially of great significance, and they have been discussed
recently within the scalar-tensor theory context  (see below).
Finally, 
for $k^2>2 $ \& $a>(3\gamma-k^2)/k$ the only late-time attracting
equilibrium point is $N$; when $a<(3\gamma-2)k/2$ these models represent
non-inflating models whereas the corresponding models inflate for
$a>(3\gamma-2)k/2$.

In the absence of an interaction term, matter scaling solutions are
represented by equilibrium points of the corresponding dynamical system. We have
shown that for simple interaction terms found in the literature, these matter scaling solutions  {\em
cannot} be represented by equilibrium points.  However, new equilibrium points arise which
represent solutions in which the energy densities of the matter and
scalar field remain a fixed proportion to one another and
obey $\gamma_\phi < \gamma$; these solutions are analogues of the matter scaling solutions in
which $\gamma_\phi=\gamma$ \cite{Wetterich1995a}.

In \cite{Amendola1999a} a large class of non-minimally coupled
scalar field models with a perfect fluid matter component were
investigated.  These models contain scalar-tensor theory models and,
in particular, Brans-Dicke theory models with a power-law potential.
On performing a conformal rescaling of the metric, the governing
equations of these models reduce to the equations for a scalar field
in general relativity with an exponential potential and an extra
coupling to the ordinary matter, and are equivalent mathematically to
the equations studies here (as was noted in \cite{Billyard1999h}).
Amendola \cite{Amendola1999a} performed a phase space analysis of
these models and obtained similar results to those obtained
here. (Since Amendola assumed $0<\gamma<2$ he obtained a wider range
of possible behaviours; however, these additional results are of lesser
interest in the context of our work. In particular, values of
$\gamma>4/3$ lead to negative values for the coupling constant a.)

The main aim in \cite{Amendola1999a} was to express the solutions back
in the original `Jordan' frame and study the cosmological consequences
of the underlying scalar-tensor theory models. In particular,
`decaying cosmological constant' solutions were considered which are
inflationary and such that the scalar field component is
asymptotically non-negligible. Models consistent with the observations
of accelerated expansion \cite{Perlmutter1999a,Riess1998a} and in which
a physically acceptable fraction of the energy-density is in the
scalar field, were found to be severely constrained by the upper
limits on the variability of the gravitational constant
\cite{Guenther1998a} and by nucleosynthesis observations.  In further
work Amendola \cite{Amendola1999c} considered a quintessential scalar
field coupled to matter with an additional radiation matter component,
and studied the effect of density perturbations on the cosmic
microwave background in these so-called `coupled quintessence' models.

\subsection{Early-time behaviour}

{}From table \ref{i_table2} we can see that the only
early-time attractors are ${\cal K}^\pm$ for certain values of $k$ and
$a$, which correspond to massless scalar field models (which are
analogues of Jacob's vacuum solutions \cite{Collins1971a}).  However,
in table \ref{i_table2} we see that for $k^2>6$ and
$a>\sqrt\frac{3}{2}(2-\gamma)$ there are no equilibrium points which
represent sources and the trajectories consequently asymptote into the
past towards a heteroclinic cycle.  In this cycle, orbits quickly
shadow the invariant set $x^2+y^2=1$ ($\Omega=0$), spend a period of
time near the saddle ${\cal K}^-$, quickly shadow the line $y=0$
($V=0$, $\Omega\neq 0$), and then spend an period of time near ${\cal
K}^+$ after which the cycle is repeated. During each cycle the orbits
pass through the inflationary portion of phase space. We stress here
that this motion is {\em not} periodic; on each successive cycle orbits
will spend a longer time near the saddles ${\cal K}^\pm$.  This
past-asymptotic qualitative behaviour, which is depicted in figure 3,
is similar to that found in \cite{Billyard1999c} within the context of
string cosmology; this is not surprising due to the conformal
relationship between these string models and the models under
investigation here \cite{Billyard1999h}.

\subsection{The case $\gamma=2$}

When $\gamma=2$, corresponding to a stiff perfect fluid, the models is
equivalent to a model containing a scalar field with an exponential
potential and a second interacting massless scalar field.  We also
note that $\gamma=2$ is a bifurcation value.  The equilibrium points
and their eigenvalues in this case are
\begin{enumerate}
\item $\displaystyle {\cal K}^+:\qquad \left(\lambda_1,\lambda_2\right) = \left(-\sqrt6 a,\sqrt{\frac{3}{2}}\left[\sqrt6 + k\right]\right). $\\
Since $a>0$ this equilibrium point is a saddle.

\item $\displaystyle {\cal K}^-:\qquad \left(\lambda_1,\lambda_2\right) = \left(\sqrt6 a, \sqrt{\frac{3}{2}} \left[\sqrt6 - k\right]\right).$\\
${\cal K}^-$ is a source for $k^2<6$ and a saddle otherwise.

\item $\displaystyle P_{\cal S}:\qquad \left(\lambda_1,\lambda_2\right) = \left(-\half\left[6-k^2\right], -\left[6-k^2-ka\right]\right).$\\
$P_{\cal S}$ is a source for $a<(6-k^2)/k$ and a saddle
otherwise.

\item $\displaystyle N:\qquad (x,y)=\left(-\sqrt\frac{6}{\Delta},
	\sqrt{\frac{a}
	{\Delta}} \right), \qquad
	\Omega=\frac{k\Delta-6}{\Delta^2},$ 
\qquad
$\displaystyle q=\frac{2\left[2k-a\right]}{\Delta},$ 
	\hspace*{4em}\\
where $\Delta\equiv k+a>0$.  Note that this solution is physical
(i.e., $\Omega\geq 0)$ either for $k^2>6$ or for $k^2<6$ and $a\geq
(6-k^2)/k$.  These models inflate
for $k^2<2$ as well as for $k^2>2$ and $a\geq2k$.  
The eigenvalues for this equilibrium point are 
\begin{eqnarray}
\lambda_\pm &=& \frac{-3a}{2\Delta}
\pm \frac{\sqrt{9a^2 
	-12a\Delta\left[k\Delta-6\right]
		}}{2\Delta},
\end{eqnarray}
and so $N$ is always a sink when it exists and is a spiral sink for
$a> [8k]^{-1}\left[27-8k^2+\sqrt{729-48k^2}\right]$ or $a<
[8k]^{-1}\left[27-8k^2-\sqrt{729-48k^2}\right]$ (and is a spiral sink
for all $k^2>6$).  Note that the scalar field acts as a perfect fluid
with an equation of state parameter given by
\be
\gamma_\phi = \frac{2}{1+\frac{1}{6}a\Delta} < \gamma.
\ee
\end{enumerate}

Note that the equilibrium point $N_2$ {\em does not exist} for
$\gamma=2$.  Table \ref{i_table2a} summarizes the stability analysis
for $\gamma=2$.  The qualitative behaviour is similar to that of the
case $\gamma\neq 2$, except that there is no region corresponding to
$3\gamma < k^2 < 6$ when $\gamma= 2$.  Note again that there exists a
heteroclinic cycle at early times for $k>\sqrt6$ (see Fig. 3); unlike
the general case $\gamma\neq2$, this heteroclinic cycle exists for all $a>0$.
\begin{table}[ht]
\begin{center}
\begin{scriptsize}
\begin{tabular}{|l||c|c||c|c|c||c|c|}
\hline
& \multicolumn{2}{|c||}{$0< k^2 < 2$} & \multicolumn{3}{|c||}{$2<k^2<6$}
&\multicolumn{2}{|c|}{$k^2>6$}\\ 
\hline 
& $a<C_1$ & $a>C_1$ & $a<C_1$ &$C_1<a<2k$ & $a>2k$&$a<2k$ 
& $a>2k$
\\ \hline 
\hline
${\cal K}^+$ 	& \multicolumn{7}{|c|}{\rule[0em]{0em}{1.2em}\raisebox{-.7em}
		[0pt]{s}} \\ 
   {\tiny(NI)}	& \multicolumn{7}{|c|}{ } \\\hline 
${\cal K}^-$ 	& \multicolumn{5}{|c||}{\rule[0em]{0em}{1.2em}
			\raisebox{-.7em}[0pt]{\bf R}}&\multicolumn{2}{|c|}
			{\raisebox{-.7em}[0pt]{s}} \\ 
   {\tiny(NI)}	&  \multicolumn{5}{|c||}{ } &\multicolumn{2}{|c|}{ }\\ \hline 
$P_{\cal S}$ 	&  \rule[0em]{0em}{1.2em} {\bf A} & s & {\bf A} 
			& \multicolumn{2}{|c||}{s} &\multicolumn{2}{|c|}
			{\raisebox{-.7em}[0pt]{\tiny DNE}} \\
	& {\tiny (I)} & {\tiny (I)}& {\tiny (NI)} 
		 &\multicolumn{2}{|c||}{{\tiny(NI)}}&\multicolumn{2}{|c|}{ } \\
		\hline
$N$ 		 & \rule[0em]{0em}{1.2em}\raisebox{-.7em}[0pt]{\tiny DNE} & {\bf A} 
			& \raisebox{-.7em}[0pt]{\tiny DNE} & {\bf A}& {\bf A}
			&{\bf A}&{\bf A}\\
   	&     	 & {\tiny (I)} &   & {{\tiny (NI)}} & {{\tiny (I)}} 
			& {{\tiny (NI)}} &{{\tiny (I)}}\\\hline
\end{tabular}
\end{scriptsize}
\end{center}
\caption[Equilibrium points for $\delta=-a\dot\phi\mu$]{{\em The
equilibrium points for the model with $\delta=-a\dot\phi\mu$ and their
stability for various values of $k$ and $a$ with $\gamma=2$.  Note
that $C_1\equiv(6-k^2)/k$.  See caption for table \ref{i_table2} for
notation.}\label{i_table2a}}
\end{table}

\begin{figure}[htp]
  \centering
   \includegraphics*[width=5in]{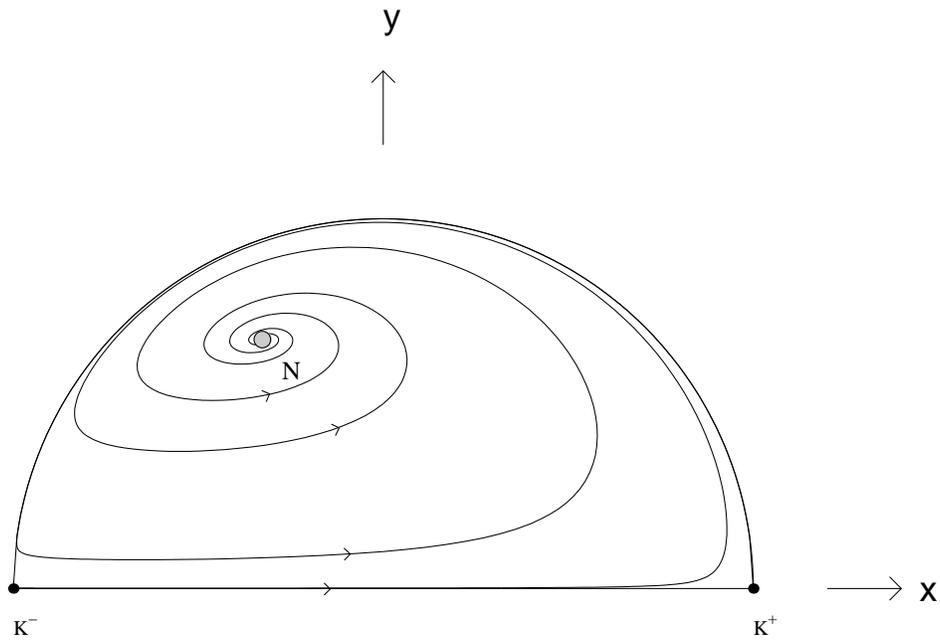}
  \caption{{\em Phase diagram of the system (\ref{i_xy_1})
when $\delta=-a\dot\phi\mu$ for the choice of parameters $\gamma=2$
and $k>\sqrt6$.  Note that the past attractor is a heteroclinic
cycle.  See also caption to figure2.}}
\end{figure}

\section{Interaction term of the form $\delta=a\mu H$\label{i_II}}

This section provides a second example to demonstrate that other types
of interaction terms can also lead to similar behaviour; i.e., that there
will be a range of parameters for which the inflationary models which
drive the matter fields to zero are {\em not} late-time attractors
and for which the trajectories exhibit an oscillatory behaviour as
they asymptote toward the late-time attracting solution.  Specifically, the
interaction term $\delta=a\mu H$ is chosen, where $a>0$.  

With this choice, equations (\ref{i_xy_1}) become
\beqn
\mainlabel{i_xy_3}
\label{i_x_3}
x' & =& -3x(1-x^2) - \sqrt{\frac{3}{2}} k y^2 + \left(\frac{3}{2}\gamma x 
	- \frac{a}{2x}\right)\left(1-x^2-y^2\right), \\
\label{i_y_3}
y' &=& \frac{3}{2}y \left[\sqrt{\frac{2}{3}} k x + 2x^2 + 
	\gamma (1-x^2 -y^2)\right].
\eeqn
For physical reasons we are not interested in
early-time behaviour, and hence the line $x=0$ will not be considered.
Consequently, a full phase-plane analysis is not possible using these
variables.  However, it is still possible to determine the equilibrium
points with $x\neq0$ for the system and determine their local
stability.

\noindent There are four equilibrium points for this system for $x\neq0$:
\begin{enumerate}
\item ${\cal K}^+$: Eigenvalues for this equilibrium point are 
\be
\left(\lambda_1,\lambda_2\right) = \left(3[2-\gamma] +a,
		\sqrt{\frac{3}{2}}\sqrt6 + 3\right)
\ee
This equilibrium point is a source.

\item ${\cal K}^-$: Eigenvalues for this equilibrium point are 
\be
\left(\lambda_1,\lambda_2\right) = \left(3[2-\gamma]+ a, 
	-\sqrt{\frac{3}{2}} +3 \right).
\ee
${\cal K}^-$ is a source for $k^2<6$ and a saddle otherwise.

\item $\displaystyle P_{\cal S}$: Eigenvalues for this equilibrium point are 
\be
\left(\lambda_1,\lambda_2\right) = \left(-\half\left[6-k^2\right],
	-\left[\Xi-k^2\gamma-k^2\right]\right),
\ee
where $\Xi\equiv 3\gamma-a$.  This point exists only for $k^2<6$, and is a source for $k^2<3\gamma$ and $a<3\gamma-k^2$ (saddle otherwise).

\item $\displaystyle N_1:\qquad (x,y)=\left(-\frac{\Xi}{\sqrt 6k},
	\sqrt{\frac{a}{3\gamma}+\frac{(2-\gamma)}{\gamma}\frac{\Xi^2}{6k^2}} 
	\right), \qquad \Omega=\frac{\Xi(k^2-\Xi)}{3\gamma k^2},$ 
\quad $\displaystyle q=-1+\half\Xi,$ 
	\hspace*{4em}\\
Note that this solution is only physical ($\Omega\geq 0)$ for
$[3\gamma-k^2]<a<3\gamma$ and $k^2<3\gamma$, and represents an
inflationary model for $a>(3\gamma-2)$ (consequently, this model will
always inflate for $k^2<2$ and can inflate for $2<k^2<3\gamma$). The
eigenvalues for this equilibrium point are
\begin{eqnarray}
\label{newsink_2}
\lambda_\pm &=& -\frac{[2a(2\Xi-k^2)+(2-\gamma)\Xi^2]}{4\gamma\Xi}
	\nonumber 
	\\ \nonumber 	&& 
\pm \frac{\sqrt{\left[2a(2\Xi-k^2)+(2-\gamma)\Xi^2\right]^2k^2 
	-8\gamma\Xi^2(k^2-\Xi)\left[2ak^2+(2-\gamma)\Xi\right]}}
	{4\gamma\Delta k},\\
\end{eqnarray}
and so $N_1$ is a sink for $a<3\gamma-\half k^2$.  Note that the scalar
field acts as a perfect fluid with an equation of state parameter given by
\be
\gamma_\phi = \frac{\gamma}{1+\frac{ak^2}{\Xi^2}} < \gamma.
\ee
\end{enumerate}

Table \ref{i_table3} lists the equilibrium points and their stability
for the ranges of $k$ and $a$.  Again, for the range $k^2<2$ and
$(3\gamma-\half k^2)<a<(3\gamma-\half k^2)$, the power-law model
$P_{\cal S}$ is no longer a sink and $N_1$ is a source.  Therefore,
solutions approach the equilibrium point which is represented by
$P_{\cal S}$ (thereby inflating) for an indefinite period of time, but
eventually evolve away.  It can be shown numerically that within this
range for $k$ and $a$, the equilibrium point $N_1$ is a spiral node
(for instance, for $k=1$ and $\gamma=4/3$, $N_1$ is a sink for
$3<a<3.5$ and a spiral sink for $3.24 \lesssim a < 3.5$); therefore
the scalar field exhibits oscillatory motion as the solutions
asymptote toward $N$.  Care must be taken in interpreting the analysis
and obtaining global results since this system is not well defined for
$x=0$ ($\dot\phi=0$) for this particular example.

\begin{table}[ht]
\begin{center}
\begin{tabular}{|l||c|c|c||c|c|c||c|}
\hline
& \multicolumn{3}{|c||}{$0< k^2 < 2$} & \multicolumn{3}{|c||}{$2<k^2<3\gamma$}
& $3\gamma < k^2 < 6$\\ 
\hline 
& $a<C_3$ & \multicolumn{2}{|c||}{$a>C_3$} & $a<C_3$ & \multicolumn{2}{|c||}{$a>C_3$} &  \\ \hline
& & $a<C_4$ & $a>C_4$ &  & $a<C_4$ & $a>C_4$ & \\ \hline 
\hline
${\cal K}^\pm$ 	& \multicolumn{7}{|c|}{\rule[0em]{0em}{1.2em}{\bf R}} \\ 
   		& \multicolumn{7}{|c|}{{\tiny(NI)} } \\ \hline 
$P_{\cal S}$ 	&  \rule[0em]{0em}{1.2em} {\bf A} & \multicolumn{2}{|c||}{s} 
		& {\bf A}  & \multicolumn{3}{|c|}{s}\\
		& {\tiny (I)} & \multicolumn{2}{|c||}{{\tiny (I)}} 
		& {\tiny (NI)} &\multicolumn{3}{|c|}{{\tiny(NI)}} \\
		\hline
$N$ 		& \rule[0em]{0em}{1.2em} & {\bf A} 
		& s &  & {\bf A}& s  & 	 \\
   		&  {\tiny DNE} & {\tiny (I)}& {\tiny (I)} 
			& {\tiny DNE}  & {\tiny (I) for}
			& {\tiny (I) for}  & {\tiny DNE} \\
   		&     	 &   &  &   & \raisebox{.7em}[0pt]{\tiny $a>3\gamma-2$}
			& \raisebox{.7em}[0pt]{\tiny $a>3\gamma-2$}  &  \\
\hline
\end{tabular}
\end{center}
\caption[Equilibrium points for $\delta=a\mu H$]{{\em The equilibrium
points for the model with $\delta=a\mu H$ and their stability for
various values of $k$ and $a$.  Note that $C_3\equiv3\gamma-k^2$ and
$C_4\equiv3\gamma-\half k^2$.  The symbol ``{\bf R}'' denotes when the
equilibrium point is a source (repellor), ``s'' for when it is a
saddle, ``{\bf A}'' for when it is a sink (attractor), and ``DNE''
when it does not exist within the particular parameter space.  The
label ``(NI)'' denotes non-inflationary models whereas ``(I)''
represents inflationary models.  For $k^2>6$ the only equilibrium
points to exist are the points ${\cal K}^\pm$; ${\cal K}^+$
is a source and ${\cal K}^-$ is a saddle (i.e., there is no sink).
Recall that the system is not well defined for $x=0$ and therefore a
global analysis is not possible.  }\label{i_table3}}
\end{table}

\section{Conclusions\label{i_discuss}}

Without an interaction term, it is known that for $k^2<2$ the global
late-time attractor for the system (\ref{i_xy_1}) is a power-law
inflationary model in which the matter is driven to zero
\cite{Copeland1997a}.  The purpose of this paper was to show that this
behaviour could be altered qualitatively with the introduction of an
interaction term. In particular, for models with an interaction term
of the form $\delta=-a\dot\phi\mu$ there are values in the parameter
space for which the equilibrium point $P_{\cal S}$, corresponding to
this particular power-law inflationary model, becomes a saddle, and so
while the models may spend an arbitrarily long period of time
inflating with $\Omega\rightarrow 0$, they eventually evolve away from
this solution.  The late-time attractors within the same parameter
space, corresponding to the new attracting equilibrium point $N_1$,
are also inflationary but with the matter field's energy density
remaining a fixed fraction of the scalar field's energy-density and
with $\gamma_\phi < \gamma$.  These are analogues of the matter
scaling solutions in which $\gamma_\phi=\gamma$.

For an appropriate parameter range, the equilibrium point $N_1$ is an
attracting focus, and hence as solutions approach this late-time
attractor the scalar field oscillates.  Although the late-time
behaviour (corresponding to $N_1$) is still inflationary, the
oscillatory behaviour provides a possible mechanism for inflation to
stop and for conventional reheating to ensue (indeed this is similar
to the mechanism for reheating in scalar field models with a potential
containing a global minimum
\cite{Linde1987a,Linde1983a,Linde1983b,Amendola1996a,Berera1996a}). To
study reheating properly more complicated physics in which the
oscillating scalar field is coupled to both fermionic and bosonic
fields needs to be included.  This contrasts with the situation for
exponential models for $k^2<2$ with no interaction term which have no
graceful exit from inflation and in which there is no conventional
reheating mechanism.

Therefore, we have shown that there are general relativistic scalar
field models with an exponential potential which evolve towards an
inflationary state in which the matter is not driven to zero and which
exhibit late-time oscillatory behaviour; these models may constitute a
first step towards a more realistic model.  There is the question of
how  physical these models are, since they correspond to relatively
large values of $a$.  In the context that the interaction
term represents energy transfer, for physical reasons it might be
expected that $a$ must be small; i.e., $a<0.1$ \cite{Wetterich1995a}
(see also \cite{Amendola1999b}).  On the other hand, in the
context of scalar-tensor theories $a$ is of order unity and can certainly
attain values large enough to produce the behaviour described above (see
Eq. (\ref{couplingfromST}); this is also the situation in
the context of string theories.

It is also of interest to study the cosmological consequences of the
`decaying cosmological constant' or `quintessential' cosmological
models, since they may be consistent with the observations of
accelerated expansion \cite{Perlmutter1999a,Riess1998a} and may lead to a
physically interesting current residual scalar field energy-density.
These issues have recently been addressed by Amendola
\cite{Amendola1999a,Amendola1999c} in the context of the conformally
related scalar-tensor theories of gravity.

\

\

\hspace*{\fill}{\bf Acknowledgments}\hspace*{\fill}

APB is supported by Dalhousie University and AAC is supported by the
Natural Sciences and Engineering Research Council of Canada,

\

\


\begin{thebibliography}{10}

\bibitem{Green1987a}
M.~B. Green, J.~H. Schwarz, and E.~Witten,
\newblock {\em Superstring Theory},
\newblock Cambridge University Press, 1987.

\bibitem{Olive1990a}
K.~A. Olive,
\newblock Phys. Rep. {\bf 190}, 307 (1990).

\bibitem{Billyard1999h}
A.~P. Billyard,
\newblock {\em The Asymptotic Behaviour of Cosmological Models Containing
  Matter and Scalar Fields},
\newblock PhD thesis, Dalhousie University, 1999.

\bibitem{Guth1981a}
A.~H. Guth,
\newblock Phys. Rev. D {\bf 23}, 347 (1981).

\bibitem{Caldwell1998a}
R.~R. Caldwell, R.~Dave, and P.~J. Steinhardt,
\newblock Phys. Rev. Lett. {\bf 80}, 1582 (1998).

\bibitem{Bahcall1999a}
N.~Bahcall, J.~P. Ostriker, S.~Perlmutter, and P.~J. Steinhardt,
\newblock Science {\bf 284}, 1481 (1999).

\bibitem{Perlmutter1999a}
S.~Perlmutter et~al.,
\newblock Astrophys. J. {\bf 517}, 565 (1999).

\bibitem{Riess1998a}
A.~G. Riess et~al.,
\newblock Astron. J. {\bf 116}, 1009 (1998).

\bibitem{Guzman1998a}
F.~S. Guzman, T.~Matos, and H.~{Villegas-Brena},
\newblock {\em Dilatonic dark matter in spiral galaxies},
\newblock astro-ph/9811143, 1998.

\bibitem{Huterer1998a}
D.~Huterer and M.~S. Turner,
\newblock {\em Revealing quintessence},
\newblock astro-ph/9808133, 1998.

\bibitem{Halliwell1987a}
J.~J. Halliwell,
\newblock Phys. Lett. B {\bf 185}, 341 (1987).

\bibitem{Burd1988a}
A.~B. Burd and J.~D. Barrow,
\newblock Nucl. Phys. B {\bf 308}, 929 (1988).

\bibitem{Lidsey1992a}
J.~E. Lidsey,
\newblock Class. Quantum Grav. {\bf 9}, 1239 (1992).

\bibitem{Aguirregabiria1993b}
J.~M. Aguirregabiria, A.~Feinstein, and J.~{Ib\'{a}\~{n}ez},
\newblock Phys. Rev. D {\bf 48}, 4662 (1993).

\bibitem{Feinstein1993a}
A.~Feinstein and J.~{Ib\'{a}\~{n}ez},
\newblock Class. Quantum Grav. {\bf 10}, 93 (1993).

\bibitem{Kitada1993a}
Y.~Kitada and M.~Maeda,
\newblock Class. Quantum Grav. {\bf 10}, 703 (1993).

\bibitem{Coley1997a}
A.~A. Coley, J.~{Ib\'{a}\~{n}ez}, and R.~J. {van den Hoogen},
\newblock J. Math. Phys {\bf 38}, 5256 (1997).

\bibitem{Wetterich1988a}
C.~Wetterich,
\newblock Nucl. Phys. B {\bf 302}, 668 (1988).

\bibitem{Wands1993a}
D.~Wands, E.~J. Copeland, and A.~R. Liddle,
\newblock Ann. N.Y. Acad. Sci. {\bf 688}, 647 (1993).

\bibitem{Ferreira1998a}
P.~G. Ferreira and M.~Joyce,
\newblock Phys. Rev. D {\bf 58}, 023503 (1998).

\bibitem{vandenHoogen1999b}
R.~J. {van den Hoogen}, A.~A. Coley, and D.~Wands,
\newblock Class. Quantum Grav. {\bf 16}, 1843 (1999).

\bibitem{Coley1998b}
A.~A. Coley, J.~{Ib\'{a}\~{n}ez}, and I.~Olasagasti,
\newblock Phys. Lett. A {\bf 250}, 75 (1998).

\bibitem{Billyard1999f}
A.~P. Billyard, A.~A. Coley, R.~J. van~den Hoogen, J.~Ibanez, and
  I.~Olasagasti,
\newblock {\em Scalar field cosmologies with barotropic matter: Models of
  {B}ianchi class {B}},
\newblock submitted to Classical and Quantum Gravity, 15 pages, gr-qc/9907053,
  1999.

\bibitem{Linde1987a}
A.~Linde,
\newblock Inflation and quantum cosmology,
\newblock in {\em 300 Years of Gravitation}, edited by S.~W. Hawking and
  W.~Israel, pages 604--630, Cambridge University Press, Cambridge, 1987.

\bibitem{Linde1983a}
A.~Linde,
\newblock Phys. Lett. B {\bf 129}, 177 (1983).

\bibitem{Linde1983b}
A.~Linde,
\newblock JETP lett. {\bf 38}, 176 (1983).

\bibitem{Amendola1996a}
L.~Amendola, C.~Baccigalupi, and F.~Occhionero,
\newblock Phys. Rev. D {\bf 54}, 4760 (1996).

\bibitem{Berera1996a}
A.~Berera,
\newblock Phys. Rev. D {\bf 54}, 2519 (1996).

\bibitem{Berera1995a}
A.~Berera,
\newblock Phys. Rev. Lett. {\bf 74}, 1912 (1995).

\bibitem{Bellini1999a}
M.~Bellini,
\newblock {\em Towards a theory of warm inflation of the universe},
\newblock gr-qc/9904072, 1999.

\bibitem{Albrecht1982b}
A.~Albrecht, P.~J. Steinhardt, and M.~S. Turner,
\newblock Phys. Rev. Lett. {\bf 48}, 1437 (1982).

\bibitem{Berera1995b}
A.~Berera,
\newblock Phys. Rev. Lett. {\bf 75}, 3218 (1995).

\bibitem{Oliveira1998a}
H.~P. de~Oliveira and R.~O. Ramos,
\newblock Phys. Rev. D {\bf 57}, 741 (1998).

\bibitem{Yokoyama1987a}
J.~Yokoyama, K.~Sato, and H.~Kodama,
\newblock Phys. Lett. B {\bf 196}, 129 (1987).

\bibitem{Yokoyama1988a}
J.~Yokoyama and K.~Maeda,
\newblock Phys. Lett. B {\bf 207}, 31 (1988).

\bibitem{Wetterich1995a}
C.~Wetterich,
\newblock Astron. Astrophys. {\bf 301}, 321 (1995).

\bibitem{Eckart1940a}
C.~Eckart,
\newblock Phys. Rev. {\bf 58}, 919 (1940).

\bibitem{Copeland1997a}
E.~J. Copeland, A.~R. Liddle, and D.~Wands,
\newblock Phys. Rev. D {\bf 57}, 4686 (1998).

\bibitem{Collins1971a}
C.~B. Collins,
\newblock Comm. Math. Phys. {\bf 23}, 137 (1971).

\bibitem{Carr1999a}
B.~Carr and A.~A. Coley,
\newblock Class. Quantum Grav. {\bf 16}, R31 (1999).

\bibitem{Amendola1999a}
L.~Amendola,
\newblock Phys. Rev. D {\bf 60}, 043501 (1999).

\bibitem{Guenther1998a}
D.~B. Guenther, L.~M. Krauss, and P.~Demarque,
\newblock Astrophys. J. {\bf 498}, 871 (1998).

\bibitem{Amendola1999c}
L.~Amendola,
\newblock {\em Perturbations in a coupled scalar field cosmology},
\newblock astro-ph/9906073, 1999.

\bibitem{Billyard1999c}
A.~P. Billyard, A.~A. Coley, and J.~Lidsey,
\newblock {\em Qualitative analysis of early universe cosmologies},
\newblock gr-qc/9907043, Accepted to J. Math. Phys. (October), 1999.

\bibitem{Amendola1999b}
L.~Amendola,
\newblock {\em Coupled quintessence},
\newblock astro-ph/9908023, 1999.

\end{thebibliography}

\end{document}